\documentstyle[12pt]{article}
\newcommand{\be}{\begin{equation}}
\newcommand{\ee}{\end{equation}}

\begin{document}
\input epsf
\begin{titlepage}
%\today          \hfill 
\begin{center}
%\hfill    Preprint No \\
\vskip .2in
{\large \bf VANISHING VIERBEIN
IN GAUGE THEORIES OF GRAVITATION}
\footnote{Preprint Uni Goettingen published August 1982}
%\footnote{Thanks text}
\vskip .50in
%alternate footnote for faculty:

\vskip .2in
A.~Jadczyk${}^\sharp$\footnote{
e-mail: ajad@physik.uni-bielefeld.de}\\
Institut f\"ur Theoretische Physik\\
der Universit\"at Goettingen\\
Bunsenstrasse 9, D 3400 G\"ottingen\\

\end{center}
\vskip .2in
\begin{abstract} 
We discuss the problem of a degenerate vierbein in the framework of gauge theories of 
gravitation. We show that a region of space-time with vanishing vierbein but smooth 
principal connection can be, in principle, detected by scattering experiments.
\end{abstract}

\noindent {\sl Author's comments:}

\noindent This paper was send for publication in February 1984. It then took almost a year to get Referee Report. Here its is:
\begin{quote}
	This paper contains a summary of some of the known aspects of gravity as a gauge theory and addresses, without substantive results, the phenomena associated with regions where the vierbein vanishes. What is new in the paper is connected with this latter question, but I find the discussion misleading and in any case not sufficiently well developed to justify publication in the paper's present form. Coordinate regions where the metric or vierbein vanishes must be treated with considerable care, as there are generally identifications of points to be taken into account. Pictures such as that of Fig. 2 are thus very misleading. The vanishing of the vierbein is not in general a coordinate-independent statement and what seems to be two points in one coordinate frame could well be seen to be just one in another. This is the case, e.g., with a uniformly accelerated frame in the x direction, where all points with $x=0$ and $\tau$ finite are to be identified.

\noindent December 17, 1984

\end{quote}
\noindent My comments on the above report - as of today September 18, 1999:
\begin{quote} 

In retrospect, I realize that the anonymous referee did not, in fact, reject the paper. He suggested that it did not justify  publication in its present form.  However, I was sufficiently discouraged by his comments to decide not to work on improving the form any further.  Also in retrospect, it seems clear that the referee missed my point: when all the information that we have is included in the metric then,  we can  try to play the game of identifying the points to get rid of the "singularity."  But within the framework I was discussing in this paper, there was also a principal connection  using extra dimensions and giving extra-information.  Gluing some space-time points together would  create discontinuity of the connection. Vierbein, in the paper, was defined as one-form with values in an associated vector bundle and its vanishing was a coordinate independent statement. 
Yet,  the referee was right on one point: the ideas of the paper could have been developed better.   Although I have shied away from the subject in the intervening years, at the time I was  suggesting that "faster-than-light teleportation" is possible but, indeed, I failed to  provide the explicit description of  a transdimensional remolecularizer working on this principle. \\
In the meantime the paper has been quoted in I. Bengtsson, "Degenerate metrics and an
empty black hole", {\sl Class. Quantum Grav.} {\bf 8} 1991, 1847-1857 (for more recent developments cf. also I. Bengtsson and T. Jacobson, 
"Degenerate metric phase boundaries", {\sl Class. Quantum Grav.} {\bf 14} 1997, 3109-3121 
\end{quote}
\end{titlepage}
\newpage
\section{Introduction}
Hanson and Regge \cite{hr} (see also D'Auria and Regge \cite{ar} ) 
suggested that a gravitational Meissner effect might exist producing torsion vortices 
accompanied by vanishing of a vierbein. This in turn may indicate a phenomenon of 
"unglueing" of a principal $0(4)$ (or $0(3, 1)$) bundle from the bundle of frames of 
the space-time manifold $M.$ The idea that a proper           
arena for gauge theories of gravitation is an external principal bundle $P$       
over $M,$ rather than the frame bundle $LM$ of $M,$ has been put forward by many authors 
(see e.g. \cite{petti,pilch,gia}). 
It is not our aim to justify this belief here. In fact what is relevant is not so much 
the choice of a bundle but        
the choice of dynamical variables and their dynamics. As long as a vierbein is thought 
of as representing a homomorphism of the two bundles,           
any distinction between $P$ and a subbundle of $LM$ does not really matter.         
 The distinction becomes important when the vierbein is interpreted as a         
linear map $\theta : \zeta^\mu \mapsto \theta^a_{\mu}\zeta^\mu $ from the tangent spaces of 
$M$ into the fibers of a vector bundle associated with $P.$ Such an interpretation 
is natural in gauge theories of the Poincar\'e group $IO(3,1)$ 
(see e.g. \cite{pilch, gia, hennig}) and 
of $O(3, 2)$ \cite{stelle}, where $\theta$ appears as a composite object. 
                                      
    The paper is organized as follows. in Sect. 3 we formulate a gauge theory of 
the Lorentz group and point out that the requirement of smoothness       
 of the Lagrangian at a degenerate vierbein is a strong selection criterion.       
 Only three terms survive the test, one of them having a "wrong parity",         
 and two others are (cosmetically improved) the standard Einstein-Hilbert        
 Lagrangian and the cosmological term. In Sect. 4 we briefly discuss the         
 structure of the gauge theory of the Poincare group, and of the $O(3, 2)$        
 gauge theory considered by Stelle and West \cite{stelle}. In Sect. 5 we analyse 
relations between these abstract gauge theories and the conventional ones         
 based on metric tensor and an affine connection. The relation can be visualized 
as follows. In  an open region where the vierbein $\theta$ is nondegenerate the 
external bundle is glued with the help of $\theta$ to the frame bundle
 and the bundle connection can be interpreted as an affine connection on $M$.
 When $\theta$ becomes degenerate the external bundle detaches from the frame      
bundle. The principal connection chooses to live in the external bundle and
 the affine connection dies. Two examples are given in Sect. 6 . The first is
 a kind of gravitational instanton considered by Hanson and Regge \cite{hr} and
 D'Auria and Regge \cite{ar}. A zero of a vierbein is accompanied by non-zero       
 torsion. The second example is an adaptation of a model discussed by Einstein and 
Rosen \cite{er}. The vierbein vanishes here on a 3-dimensional           
 "bridge" connecting two mirror copies of the exterior Schwarzschild universe. 
It is interesting to notice that the principal connection continues
 smoothly over the bridge and need not be regularized as in the first example. 
The Einstein--Rosen bridge is therefore a torsion-free regular solution
of vacuum field equations of $0(3, 1)$ gauge theory.

It was pointed out in \cite{ar} that "the vanishing vierbein at some point is not
 a disastrous feature of theory". It is one of the aims of the present note to
 point out that with an appropriate dynamics vanishing vierbein in a whole
 region need not be a disaster either. It is shown in Sect. 7 that such a "dead"
 region can have observable effects seen from outside, and that it introdu\-
 ces statistical elements (that is "freedom of choice")
 already on a classical level.  
\section{Notation}
Let $(P,\pi, M, G)$ be a principal bundle, let $\rho$ be a representation of $G$ on a 
vector space $F,$ and let $\rho^\prime$ denote the derived representation of the Lie 
algebra $Lie(G)$ of $G$. For every $h\in Lie(G)$ let ${\tilde h}$ denote the fundamental 
vector field on $P$ generated by $h.$ An $F$-valued $p$-form, $p\leq dim(M)$, $\phi$ 
on $P$ is called tensorial of type $\rho$ if $i({\tilde h})\phi = 0$ and ${\cal L}_{\tilde h}
\phi =
 -\rho^\prime (h)$  for all $h\in Lie(G)$ (see \cite[p. 75]{koba} ). 
Let $Q(\rho) = P\times_G F$  be the vector bundle associated                      
with $P$ via the representation $\rho$ ( \cite[p. 55]{koba} ). One can identify 
$F$-valued tensorial $p$-forms of type $\rho$ on $P$ with $Q(\rho)$-valued $p$-forms on $M$ 
(see \cite[p. 76]{koba}, \cite[Ch. VII. 1]{dieu}, \cite[Ch XX. 5]{greub}). This 
identification is extensively used in the literature and we shall use it too
without further references. In particular the curvature 2-form
$\Omega=D\omega$ of a principal connection $\omega$ on $P$ can be thought of as a 2-form on
$M$ with values in $Q(Ad)$, where $Ad$ denotes the adjoint representation of $G$ 
on $Lie(G).$
\section{$O(3, 1)$ gauge theory}
Let $(P, \pi, M, O(3, 1))$ be a principal bundle over a $4$-dimensional base manifold $M.$ 
The fibers of $P$ are to be thought of as being a priori completely detached from 
fibers of the frame bundle of $M.$ Let $\rho_0$ denote the natural representation of 
$O(3, 1)$ on 
${\bf R}^4$ . The dynamical variables of a generalized Einstein--Cartan theory are: 
a principal connection $\omega$ on $P$, and a  $Q(\rho_0)$-valued $1$-form $\theta$ 
on  $M.$ In a "pure gauge theory"
\footnote{By a pure gauge theory we mean a gauge field theory where the 
primitive fields have no direct connection to space-time geometry. Such a                     
theory should "interprete itself" (see Ref. \cite{ajad82}. }  
a Lagrangian $4$-form should be constructed out of $\omega$ and $\theta$ alone. 
We demand that the action should be a smooth (and thus nonsingular) function of
field configuration variables 
$(\omega, \theta ).$ In particular ${\cal L}$ should be continuous at $\theta=0.$
 Among geometrical objects at our disposal we find only six candidates which have this property: 
\begin{eqnarray}
{\cal L}_1 &=& \epsilon_{abcd}\; \theta^a\wedge\theta^b\wedge\theta^c\wedge\theta^d\\
{\cal L}_2 &=& \epsilon_{abcd}\; \theta^a\wedge\theta^b\wedge\Omega^{cd}\\
{\cal L}_3 &=& \theta^a\wedge\theta^b\wedge\Omega_{ab}\\
{\cal L}_4 &=& D\theta^a\wedge D\theta_a\\
{\cal L}_5 &=& \epsilon_{abcd}\; \Omega^{ab}\wedge\Omega^{cd}\\
{\cal L}_6 &=& \Omega_{ab}\wedge\Omega^{ab}\end{eqnarray}

Variations of ${\cal L}_6$ and ${\cal L}_5$ are exact $4$-forms and do not 
contribute to classical field equations. Owing to the first Bianchi identity 
${\cal L}_4$ differs from 
${\cal L}_3$ by an exact form only. ${\cal L}_3$ has internal parity different from
that of ${\cal L}_2$ and ${\cal L}_1$, and should not be combined with them unless 
$P$ is reduced to $SO(3, 1)$, which would imply dynamically preferred orientation of $M$
- a viable possibility.\footnote{Another possibility is to introduce an extra pseudo-scalar field 
$\psi$ and to replace ${\cal L}_3$ with $\psi{\cal L}_3$ (see \cite{nelson}). } 
Introducing arbitrary coupling constants $\lambda_1 , \lambda_2 , \lambda_3 $, Euler-Lagrange equations for 
\be
{\cal L}=(\lambda_1/4!){\cal L}_1 + 
(\lambda_2/2){\cal L}_2 + (\lambda_3/2){\cal L}_3
\label{lag}
\ee 
are:
\be
\epsilon_{abcd}\theta^b\wedge((\lambda_0/3!)\theta^c\wedge\theta^d+\lambda_1\Omega^{cd}
)+\lambda_2\theta_b\wedge\Omega^{ab} = t_a,
\label{ta}
\ee

\be
\lambda_1\epsilon_{abcd} D\theta^c\wedge\theta^d+(\lambda_2/2)(D\theta_a\wedge\theta_b
-D\theta_b\wedge\theta_a)=s_{ab},
\label{sab}
\ee
where $t_a$ and $s_{ab}$ are $3$-forms representing the sources: 
energy-momentum and spin. It is important to notice that field equations 
(\ref{ta}) and (\ref{sab}) make sense for all smooth configurations $(\omega,\theta)$,
 in particular for those with degenerate $\theta$-s. However, 
the predictive power of these equations falls down with the rank of $\theta$.                                                     
                                                                                             
\section{Gauge theories of $IO(3, 1)$ and $O(3,2)$}                                                    
                                                                                             
  Replacing $O(3, 1)$ with $IO(3, 1)$ one gets a gauge theory of the Poincar\'e group. 
The two important representations of $IO(3, 1)$ on ${\bf R}^4$ are 
$\rho_0(a,\Lambda):x\mapsto \Lambda x ,$ and $\rho(a,\Lambda):x\mapsto \Lambda x + a.$ 
The dynamical variables of an $IO(3, 1)$ gauge theory are (cf. \cite{pilch,gia}): 
a principal connection $\omega$ on                          
$(P,\pi, M, IO(3, 1))$, and a $Q(\rho)$-valued $0$-form $\phi$ on $M ,$ $Q(\rho)$ being the
affine bundle over $M$ associated to $\rho .$ 
The admissible Lagrangians are those of the $0(3, 1 )$ gauge theory but now $\theta$ is not 
a primitive field -- it is defined by $\theta=D\Phi$. Given a field configuration 
$( \omega,\Phi )$ one can always adapt a gauge (that is to use the translational freedom and choose a cross section of the principal $IO(3,1)$ bundle) in such a way that $\Phi^a \equiv 0$ -- thus, effectively, eliminating the field $\Phi$ from the theory. In this gauge the soldering form $\theta$ , defined above as        
$\theta = D\Phi$,  coincides with the translational part $\omega^a$ of the connection $\omega .$ This closes our discussion of the Poincar\'e group gauge theory: after gauge fixing it effectively reduces to Lorentz group gauge theory.                                                  

Another viable possibility is a gauge theory of $0(3, 2)$, with five dimensional fibres, as discussed  in \cite{stelle}. The  dynamical variables
of this theory are: an $0(3, 2)$ -- connection $\omega$ and a $Q(\rho)$ -valued $0$-form 
$y,$ where $\rho$ denotes the natural representation of $O(3, 2)$ on ${\bf R}^5 .$ 
The Lagrangian 4-form is \cite{stelle}                                                                         
\be
{\cal L} =\epsilon_{ABCDE}\ y^A\wedge\Omega^{BC}\wedge\Omega^{DE},\; A,B,\ldots,E=0,1,2,3,5
\ee
together with a constraint
\be
y^A y_A=const .
\ee             
Given a configuration $(\omega,y)$ one can always adapt gauge in such a way 
that $y^5=0.$ The (generalized) vierbein $\theta$ can be defined by imposing the condition                                           
\be
\theta^a=(Dy)^a ,\; a=0,1,2,3 
\ee
in the adapted gauge.                                                                        
                                                                                                  
\section{Relation to metric-affine theories}                                                            
                                                                                                  
 Let $(\omega,\theta)$  be a field configuration of an $O(3, 1)$ gauge theory as discussed 
in Sect. 3. A point $p\in M$ is said to be a critical point of $\theta$ if 
$\det(\theta^a_\mu (p))  = 0,$ otherwise $p$ is called a regular point. 
The set $M_\theta$  of all regular points of a smooth $\theta$ is an open subset of $M .$ 
When $M_\theta \neq M$ then $\theta$ is called degenerate.                                                                                       
 A linear space-time connection $\nabla$ in $TM$ (sometimes also called an affine connection on $M$; notice that we admit torsion here)
 is said to be {\sl compatible} with a given field configuration $(\omega,\theta)$ if                         
\be
\theta(\nabla_X Y) = i(X) D i(Y)\theta
\label{comp}
\ee                                                                                           
  i. e. if $\theta$ is parallel with respect to $(\omega,\nabla)$ when considered as a 
cross section of the fiber product $Q(\rho_0)\times T^\star M .$ We have then                                         
\be
\theta(\nabla_X Y -\nabla_Y X- [X,Y]) = (D\theta)(X,Y)
\label{tor}
\ee 
and so the $Q(\rho_0)$ -- valued $2$-form $\Theta=D\theta$ may be considered as a 
generalized torsion.                                                                                

 Given a configuration $(\omega,\theta)$ there exists a unique $\nabla$ on
$M_\theta$, $M_\theta$ being the open submanifold of $M$ consisting of all regular 
points of the vierbein, compatible  with $(\omega,\theta).$ Outside of $M_\theta$ -- on the
set of critical space--time points -- the affine 
connection $\nabla$ will not, in general, exist. To see this observe that at 
regular point we have, as a consequence of Eq. (\ref{comp}):
\be
\Gamma^\sigma_{\mu\rho}=\theta^{{-1}\phantom{a}\sigma}_{\phantom{-1}a} (\partial_\mu \theta^a_\nu+
\omega_{\mu\phantom{a} b}^{\phantom{\mu}a} {\theta^b_\nu)}
\label{gam}
\ee
where $\Gamma^\sigma_{\mu\nu}$ are the coefficients of $\nabla$ in a coordinate 
system $x^\mu .$                 
From Eq. (\ref{gam}) we get

\be
\Gamma_\mu\equiv\Gamma^\sigma_{\mu\sigma}=\partial_\mu \ln \vert \det \theta \vert
\label{gamss}
\ee
so that the part of $\nabla$ which is responsible for the parallel transport
 of the length scale depends on the vierbein only (and not on the connection $\omega$).  When $\theta$ becomes degenerate 
then $\ln( \det \theta)$, and therefore also $\Gamma_\mu$, diverge. If $\theta$ is 
identically zero then the argument does not apply, and inside such a region any affine 
connection is compatible with such a configuration.        
                                       
   For every configuration $(\omega,\theta)$ one defines a covariant "metric" tensor 
$g_{\mu\nu}=\eta_{ab}\theta^a_\mu\theta^b_\nu ,$ where $\eta=diag(-1,+1,+1,+1)$ is the 
diagonal constant matrix.
The induced scalar product in $T_p M$ is nondegenerate if and only if $p$ is regular. 
On $M_\theta$ we have then $\nabla_\mu g_{\nu\sigma} =0$, where $\nabla$ is a unique 
affine connection compatible with $(\omega,\theta)$. The standard Einstein--Hilbert                   
 Lagrangian density ${\cal L}_{EH}$ of $(g,\theta)$ is                                                           
\be
{\cal L}_{EH}=(\lambda_1+\lambda_2 R)\vert \det (g)\vert^{1/2}
\label{eh}
\ee
where $R$ is the scalar curvature of $( \nabla, g)$. One easily finds that                             
\be
{\cal L}_{EH}= sgn(\det \theta)((\lambda_1/4!){\cal L}_1+(\lambda_2/2!){\cal L}_2)_{0123}
\label{ehsgn}
\ee
 where $\Phi_{0123}$  denotes the $(0,1,2,3)$ -- component of a $4$-form $\Phi$.                                
                                                                                              
 The identity (\ref{ehsgn}) holds on $M_\theta$ . Outside of this region the left hand side 
of Eq. (\ref{ehsgn}) is not defined. The right hand side is "almost" defined - if not for the $sgn(\det \theta)$. It is thus clear why taking (\ref{lag}) for the Lagrangian four--form is a better choice.                                                                 
                                                                                              
\section{Two examples of configurations with degenerate vierbein}                                    
                                                                                              
{\bf Example 1.}: Hanson and Regge \cite{hr} (see also \cite{ar}) consider an $O(4)$-- thus Euclidean -- version of
 Einstein-Cartan gauge theory. The base manifold $M$ is ${\bf R}^4$  and the vierbein 
$(\theta^a_\mu )$ is defined by                                                             
\be
\theta^0=2x^\mu dx^\mu ,
\ee
\be
\theta^1=x_0dx^1-x_1dx^0+x_2dx^3-x_3dx^2\; (cyclic).
\ee
 It is nondegenerate everywhere except at the origin where it vanishes.

The unique torsion-free connection form (compatible with $\theta$) $\omega$ given in ${\bf R}^4\setminus \{ 0 \}$ by
\be
\omega^{i0}=\omega^{jk}=-\rho^{-2}\theta^i\;  (cyclic)
\label{hr}
\ee
 is flat and singular at $x=0.$   A regularization
\be
\omega\mapsto \omega^{i0}=\omega^{jk}=-\phi(\rho^2)\theta^i,\; \rho^2=\Sigma_\mu (x^\mu)^2,
\ee
results in a non-zero torsion
\be
D\theta^i=(-\phi+\rho^{-2})(\theta^0\wedge\theta^1+\epsilon^{ijk}\theta_j\wedge\theta_k .
\ee
Thus we get a globally defined non-singular connection in an external $O(4)$ bundle. 
It induces space-time affine connection everywhere except at the origin. The induced
connection has non-vanishing torsion. 

\noindent{\bf  Example 2.}: Einstein and Rosen \cite{er} considered two examples of solutions of (modified)
 vacuum field equations of general relativity with a degenerate metric.
Let us show how both examples can be easily reformulated in terms of an $O(3, 1)$ gauge theory. 
The first model discussed in \cite{er} describes a uniformly accelerated                        
frame in a flat Minkowski space. The second, described below, brings
 similar features with a non-flat connection. One takes here 
$M={\bf R}^2\times S^2$                  
 with coordinates $x^0 =t,x^1 =u, x^2 = \psi, x^3 = \phi$, and the vierbein is given by                                    
\begin{eqnarray}
\theta^0 &=& u(u^2+2m)^{-1/2}dt,\nonumber \\
\theta^1 &=& 2(u^2+2m)^{1/2}du,\nonumber \\
\theta^2 &=& (u^2+2m)d\psi,\nonumber \\
\theta^3 &=& (u^2+2m)\sin(\psi) d\phi .\nonumber 
\end{eqnarray}
 It degenerates at $u=0$ (the "bridge"). The connection form can be                             
 represented by
\begin{eqnarray}
\omega^1_{\phantom{1} 0}&=&mu^{-1}(u^2+2m)^{-3/2}\theta^0,\nonumber \\
\omega^2_{\phantom{2} 1}&=&u(u^2+2m)^{-3/2}\theta^2,\nonumber \\
\omega^3_{\phantom{3} 1}&=&u(u^2+2m)^{-3/2}\theta^3,\nonumber \\
\omega^3_{\phantom{4} 2}&=&\cot( \psi) (u^2+2m)^{-1}\theta^3,\nonumber 
\end{eqnarray}                                                                                          
 At first sight it seems that $\omega$ is singular on the bridge in an analogy to (\ref{hr}). 
However, since the vierbein is degenerate, the $1$--forms $\theta^a$ do not form
a basis at $u=0$ and therefore not every $1$--form can be represented in terms of $\theta$--s. 
 In fact, we have
\be
\omega^1_{\phantom 0}=m(u^2+2m)^{2}dt.  \ee
Torsion vanishes here, but $\omega$ is non--flat. This configuration $(\omega,\theta)$ is a 
smooth solution of the vacuum (that is with right hand sides vanishing) field equations (\ref{ta}), (\ref{sab}) with
$\lambda_1=\lambda_3=0.$                                                                                              
\section{ Physical effects of a vanishing vierbein}
We have seen that certain Lagrangians are smooth functions of configurations 
$(\omega,\theta)$ even for degenerate $\theta$--s, The requirement of smoothness is a 
strong selection criterion. In particular it rules out quadratic terms essential in 
theories with propagating torsion \cite{hehl}. It has to be understood whether a
 singularity at 
$\det \theta=0$ stabilizes a theory. Stability properties of metric theories of 
gravitation have been discussed by many authors \cite{witten}. Theories admitting vanishing vierbein need, however, special considerations. Hanson and Regge \cite{hr} speculated about a possibility of having a "torsion foam" -- a region with vanishing vierbein and well                  
 defined principal connection. No dynamical mechanism which would make  such 
configurations stable is known. Nevertheless it can be interesting to                      
 look for a possible method of detecting such a phenomenon. Here we discuss 
motion of a test particle which meets on its way a domain with vanishing vierbein.    
                                                                           
    Consider scattering of spinless particles on a double-cone shaped region where the vierbein 
vanishes (see Fig. 1). Equations of motion for a non-spinning                          
test particle are (see \cite{ajad83})
\be
\frac{d\pi^a}{dt}+\omega_{\mu\phantom{a}b}^{\phantom{\mu}a} \frac{dx^\mu}{dt}\pi^b =0,
\label{eom1}
\ee
\be
\frac{dx^\mu}{dt}(\pi^a\theta^b_\mu-\pi^b\theta^a_\mu)=0.
\label{eom2}
\ee
Notice that four-momentum $\pi^a$ and four-velocity $\frac{dx^\mu}{dt}$ are a priori
considered as independent variables. It is only through equation (\ref{eom2}) that
momentum becomes co-linear with velocity, but this co-linearity needs to obeyed at
regular space-time only. Not inside of the vanishing vierbein domain.

In our case, in regions $(I)$ and $(III)$, where $\theta$ is invertible, momentum must be
 proportional to the velocity, as it follows from Eq. (\ref{eom2}). In consequence
trajectories
$\gamma_I$ and $\gamma_{III}$ are geodesics. In region $(II)$, where $\theta$
degenerates, momentum and velocity may be totally decoupled. Let
$(x_{ini}, \pi_{ini})$ (resp. $(x_{fin},\pi_{fin})$) denote the position and momentum 
of the particle when it enters (resp. leaves) vanishing--vierbein region $(II)$. For given 
$x_{ini},\pi_{ini},x_{fin},\pi_{fin}$ denote by $C(x_{ini},\pi_{ini},x_{fin},\pi_{fin})$
 the collection of all continuous paths
$\gamma:t\mapsto \gamma(t)$
with the following property:\ {\sl $\pi$ when paralelly transported along $\gamma$ from $x_{ini}$
to $x_{fin}$ coincides with $\pi_{fin}$ }. Observe that a past trajectory $\gamma_I$, 
and the future trajectory $\gamma_{III}$ are uniquely determined by $x_{ini},\pi_{ini}$ and $x_{fin},\pi_{fin}$ respectively. Observe also that two trajectories $\gamma_{II}$ and
${\gamma_{II}}\prime$ both belong                                                 
$C(x_{ini},\pi_{ini},x_{fin},\pi_{fin})$ if and only if $\pi_i$  is an eigenvector 
belonging to the eigenvalue zero of the flux of the curvature operator through the 
surface enclosed between $\gamma_{II}$ and
$\gamma_{II}\prime$ .\footnote{For a non-Abelian Yang-Mills field as it is in our case 
a generalized flux should be used.}  
Let  $\rho(x_{ini},\pi_{ini};x_{fin},\pi_{fin})$ be the number of elements (measure) 
of $C(x_{ini},\pi_{ini},x_{fin},\pi_{fin})$.  Would $\det \theta \neq  0$ everywhere inside 
region $(II)$ then 
$\rho(x_{ini},\pi_{ini};x_{fin},\pi_{fin}) = A\delta(x_{fin}- x_{fin0})
\delta(\pi_{fin}-\pi_{fin_0})$, 
where $x_{fin0}$ and $\pi_{fin0}$
are uniquely determined by the initial data and the geometry.
If $\theta =0$ and $\Omega=0$  in region $(II)$ then 
$\rho(x_{ini},\pi_{ini};x_{fin},\pi_{fin})=\rho_1(x_{ini},\pi_{ini};x_{fin})\delta(\pi_{fin}-
\pi_{fin0})$ where $\pi_{fin0}$ is uniquely determined by the initial data. If so, then any observed non--zero dispersion of $\rho$ implies degeneracy of $\theta$ in region $(II)$ and reflects curvature effects in this region. Can some statistical effects that are normally
attributed to quantum fluctuations be accounted for through the above mechanism?

The picture shown in Fig. 1 can be misleading in two ways. First, the time of emerging of
 the probe from region $(II)$ is random. Second, a path of the particle should be continuous
but need not be differentiable.\\
Fig. 2 shows a possible world line of a particle meeting on its way a "bubble of
 vanishing vierbein" in otherwise flat Minkowski space.                                             

\begin{figure}
\caption{A possible trajectory of a test particle through region III with vanishing vierbein}
\end{figure}
\begin{figure}
\caption{Free particle meets on its way (event $A$) a bubble ${\cal O}$ of vanishing 
vierbein. Inside the bubble the particle has no compass to orient                           
 itself; its motion is random. Since the particle entered ${\cal O}$ with a future 
pointing momentum it gets reflected from the lower walls of the                          
 diamond and undergoes a succession of chaotic motions until it "chooses" to reach the upper cone. To an external observer the particle is "teleported in no time at all" from A to B.}
\end{figure}
{\bf Acknowledgements}\\Thanks are due to H.J. Borchers, H. Goenner, F. M\"uller--Hoisen
and H. Reeh  for discussion

\newpage

\begin{thebibliography}{99}                                                      
                                                                                      
 \bibitem{hr} A. J, Hanson and T. Regge: {\em Torsion and quantum gravity}, Proc. 
Integrative Conf. on Group Theory and Mathematical Physics, University 
of Texas at Austin (1978), Lecture Notes in Physics, 94, Springer--Verlag, 
Berlin-Heidelberg-New York, 1979                                         
\bibitem{ar} B. D'Auria and T. Regge, {\sl Nucl. Phys.} {\bf B195} (1982), 308-324                          
\bibitem{petti} R. J. Petti, {\sl Gen.Rel.Grav.}, {\bf 7} (1976) 869-883                                    
\bibitem{pilch} K. A. Pilch, {\sl Lett. Math. Phys.} {\bf 4} (1980), 49-51                                   
\bibitem{gia} B. Giachetti, B. Piicci and E. Sorace, {\sl Lett. Math. Phys.}, 
{\bf 5} (1981), 85-91
\bibitem{hennig} J. Hennig and,E, Nitsch, {\sl Gen. Rel. Grav.}, {\bf 13} (1981) 947-962                     
\bibitem{mh} F. M\"uller-Hoissen and J, Nitsch, {\sl Teleparallelism -- A Viable theory of 
Gravity?} , preprint.                                                       
\bibitem{stelle} K. S. Stelle and P. C. West, {\sl Phys. Rev.} {\bf D21} (1980) 1466-1488                    
\bibitem{er} A. Einstein and N. Rosen, {\sl Phys. Rev.} {\bf 48} (1935) 73-77                            
\bibitem{koba} N. Kobayashi and K. Nomizu, {\sl Foundations of Differential Geometry}, 
Vol. I., Interscience Pubs., New York, 1963                                     
\bibitem{greub} W. Greub, S. Halperin, and B. Vanstone, {\sl Connections, Curvature, 
and Cohomology}, Vol. II., Academic Press, New York, 1972                         
\bibitem{dieu} J. Dieudonne, {\sl Treatise on Analysis}, Vol. IV, Academic Press, 
New York, 1974                                                                   
\bibitem{ajad82} A. Jadczyk, {\sl On pure gauge theories}, talk at the Conf. on 
Gauge Theories and Spinor Structures, University of Clausthal, July 1982               
\bibitem{nelson} P. C. Nelson, {\sl Phys. Lett.}, {\bf 79 A} (1980), 285-287  
\bibitem{hehl} F.W. Hehl, Y. Ne'eman, J. Nitsch and P. von der Heyde, 
{\sl Phys. Lett.}, {\bf 78 B} (1978) 102-106                                                 
\bibitem{witten} E. Witten, {\sl Commun. Math. Phys.}, {\bf 80} (1981) 381-402; 
L F Abbott and S. Deser, {\sl Nucl. Phys.} {\bf B l95} (1982) 76-96;
A. Lapedes and E. Mottola, {\sl Phys. Lett.} {\bf 110 B} (1982) 114-116;
T. Parker and C. H. Taubes, {\sl Commun. Math. Phys.}, {\bf 84} (1982), 223-238                                                                          
\bibitem{ajad83} A. Jadczyk, {\sl Ann. Inst. H. Poincare}, {\bf A38} (1983), 
99          
\end{thebibliography}
\end{document}